# Two-dimensional solid state gaseous detector based on $^{10}$B layer for thermal and cold neutrons

**S Potashev, Yu Burmistrov, A Drachev, S Karaevsky, E Konobeevski and S Zuyev**

Institute for Nuclear Research, Russian Academy of Sciences, 60-letiya Oktyabrya prospekt 7a, Moscow, 117312, Russia

E-mail: potashev@inr.ru

**Abstract.** Two-dimensional solid state gaseous detector for thermal and cold neutrons is created. The detector has active area of 128 x 128 mm$^2$, $^{10}$B neutron converter, and gas chamber with thin windows. The resistive charge-division readout is applied to determine the neutron position. The detector was tested using W-Be photoneutron source at the Institute for Nuclear Research, Moscow. The detector efficiency is estimated as ~4% at neutron wavelength $\lambda$ = 1.82 Å and 8% at $\lambda$ = 8 Å. The efficiency of the background detection was less than 10$^{-5}$ of that for thermal neutrons. The resulting pulse height resolution and the spatial resolution are estimated as ~15% and ~2.5 mm, respectively.

## 1. Introduction
A thermal and cold neutron positional sensitive detectors are key devices of small angle neutron scattering (SANS) setup for studying structure and sizes of polycrystalline object in nanotechnology and biology [1, 2]. Such detectors are also used as neutron flux monitors [3]. Recently, SANS has been applied for an investigation of charging and discharging cycle of Li-ion battery *in situ* [4].

Usually, a detectors based on $^3$He gas [5] operate at high pressure which require to use a thick entrance window (~1 cm), as the leakage of this expensive gas leads to an efficiency decrease. But a neutron flux reduction in thick window does not permit to apply detector at wavelengths more than 8 Å. An alternative detector with a separation of functions of neutron converting to charged particles in $^{10}$B layer and their further detecting in the ion chamber solves not only the problem of the gas mixture stability but permits to use cheap gases under standard conditions. The localization of an interaction point of neutron in the $^{10}$B layer plane gives a possibility to obtain the good spatial and time resolution. The detector with a thin entrance window, in contrast with a gas filled one under extreme pressure, could be used in experiments with cold neutrons.

Recently, a prototype of such detector based on three 1.3 μm layer of $^{10}$B [6] has been created. A detector with six layers of 1.4 μm $^{10}$B is proposed [7]. Its efficiency is estimated as 21% at $\lambda$ = 1.82 Å. An advantage of our detector is the thin entrance window (3mm) and a long time operation which is provided by the protective semiconducting polymer layer on the boron-aluminium cathode. To avoid a mutual diffusion the boron layer is separated from aluminium by a polyimide layer.

## 2. Operating principle of detector
A thermal neutron is captured by $^{10}$B nucleus forming excited $^{11}$B* nucleus. Then, $^{11}$B* nucleus decays emitting $^4$He and $^7$Li according to following nuclear reactions:

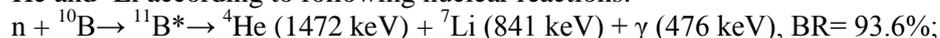
n + $^{10}$B → $^{11}$B* → $^4$He (1472 keV) + $^7$Li (841 keV) + γ (476 keV), BR= 93.6%;

$$n + {}^{10}B \rightarrow {}^{11}B^* \rightarrow {}^{4}He\ (1776\ keV) + {}^{7}Li\ (1013\ keV),\ BR = 6.4\%.$$

Sum of cross section of both reactions up to energy of 1 keV can be estimated by the formula:

$$\sigma = \sigma_0 \left(\frac{E_0}{E}\right)^{1/2} = \sigma_0 \left(\frac{\lambda}{\lambda_0}\right)^{1/2}, \qquad (1)$$

where $\sigma_0 = 3837$ b, $\lambda_0 = 1.82$ Å and $E_0 = 0.025$ eV.

The $^4$He and $^7$Li nuclei are detected in positional sensitive gaseous chamber. The chamber front cathode is founded on the glass disk of 2 mm thickness. Despite the fact that boron is semiconductor his surface quality is not suited to be used as the chamber electrode as was proposed in [6]. Therefore, the front cathode contains 3 μm $^{10}$B layer coated by a polyimid layer and 0.1 μm aluminum layer coated by protective semiconducting polymer upon the glass disk.

## 3. Multiwire and multipad ion chamber

A design of multiwire and multipad ion gas chamber is shown in figure 1. A housing of the detector consist of front and rare covers of 20 mm duralumin and a side cylindrical wall of stainless steel. The housing is sealed hermetically with a vacuum rubber. Each cover has a 3 mm entrance window for neutrons. There are placed inside the housing the following elements:

- 2-mm glass disk with 3 μm $^{10}$B layer and 0.1 μm aluminium layer. The disk serves as a front cathode.
- Anode of 64 parallel 20 μm wires of tungsten rhenium coated by gold with 2 mm spacing.
- Rare cathode of 1 mm fiberglass with the 63 copper insulated pads of 2 mm width.

A distance between the anode and each cathode is equal to 2 mm. Wires and pads are placed perpendicular to one other. The electrode assembly is enclosed in fluoroplastic housing.

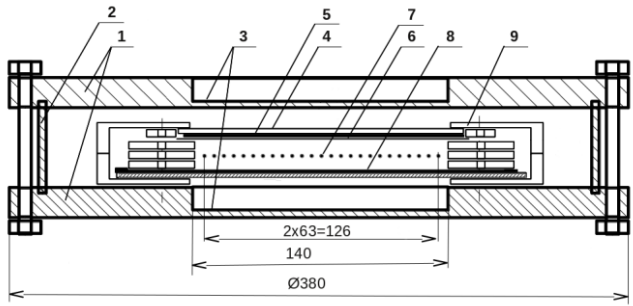

**Figure 1**. A design of detector:
1 – front and rare housing covers;
2 – cylindrical side wall of housing;
3 – windows;
4 – glass disk;
5 – boron layer;
6 – aluminum layer;
7 – wire anode of $X$ coordinate;
8 – pad rare cathode of $Y$ coordinate;
9 – fluoroplastic housing of detection assembly.

A both anode wires and rare cathode pads are connected in series to each other through a 20 Ω resistor. The both ends of the anode resistor chain through the high voltage capacitors are connected to the $X_1$ and $X_2$ preamplifiers for measuring $X$ coordinate. The both ends of the rare cathode resistor chain are connected to the $Y_1$ and $Y_2$ preamplifiers for measuring $Y$ coordinate. As soon as the pulse height from any $Y_1$ or $Y_2$ preamplifier output exceeds a threshold of the discriminator (CAEN C808) the trigger occurs. The trigger starts a conversion in the amplitude-to-digital converter (ORTEC AD811). Coordinates $X$ and $Y$ are determined from the $X_1$, $X_2$, $Y_1$ and $Y_2$ pulse heights. Accumulated data from ADC is written on computer disk. The front cathode is connected to preamplifier and discriminator. A discriminator pulse could be used as a trigger and stop signal in a time-of-flight measurement. The positive anode voltage is set from 620 V to 920 V for the gas mixture Ar+25%$CO_2$ + 0.3%$CF_3Br$ under pressure of 1.05 bars. An internal volume of detector is equal to 3.5 liters.

## 4. Experimental data analysis

Detector was tested using a neutron source. Tungsten beryllium photoneutron source (IN-LUE) was created on the base of industrial 8 MeV electron linac LUE-8, tungsten electron-gamma convertor, photoneutron beryllium target and polyethylene moderator of fast neutrons. The maximal flux density of thermal neutrons in the center of the source is about of $10^7$ cm$^{-2}$ s$^{-1}$. Detector is located at a distance of 6 m from the source at an angle of 60º relative to the electron beam axis.

The two-dimensional diagram of pulse heights $X_1$ and $X_2$ at the anode voltage 700 V is shown in figure 2. An event is registered if any pulse height $Y_1$ or $Y_2$ exceeds a preset threshold. It means that the secondary charged particle has produced ionization in both gaps of the chamber.

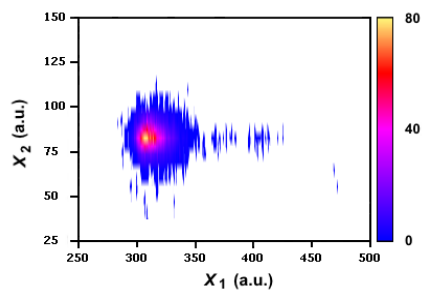
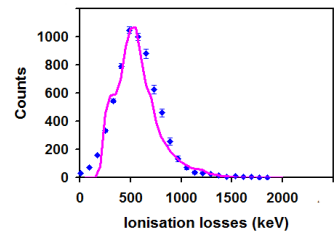

**Figure 2.** Pulse heights $X_1$ vs $X_2$ at 700 V.

**Figure 3.** Sum normalized pulse height (points) and simulated ionization loss (solid line) spectra at 700 V.

The normalized pulse height sum spectrum is shown in figure 3. The pulse height resolution as a full width at half maximum (FWHM) is equal 15% at 700 V.

The coordinate spectrum is shown in figure 4. The periodical structure in the spectrum at 700 V can be explained by a variation of electrical field near and between wires. A fact of observation of the structure leads to an estimation of spatial resolution ~ 2.5 mm.

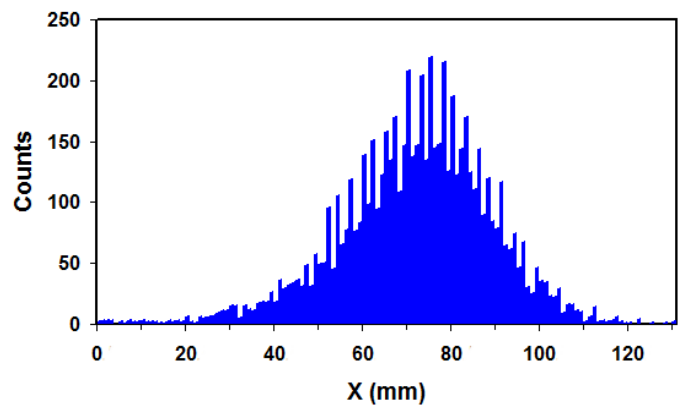

**Figure 4.** Coordinate spectrum at 700 V.

One can see from figure 5 that the form of pulse heights $X_1$ and $X_2$ at 800 V becomes broader. A height pulse corresponds to ionization loss in the gas gap. The $X_2$ height pulse and the simulated ionization loss spectrum are shown in figure 6. Comparing the measured and simulated spectra we can estimate the energy threshold as ~0.2 MeV and maximum position as ~0.55 MeV.

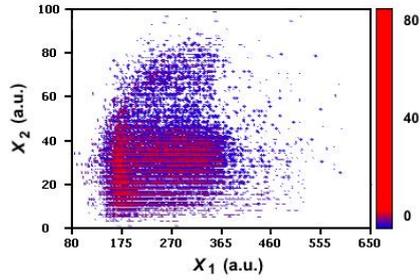 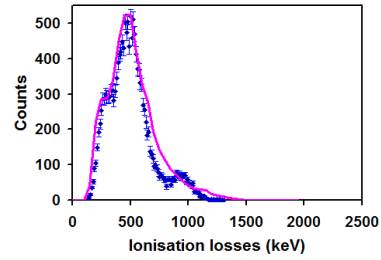

**Figure 5.** Pulse heights $X_1$ vs $X_2$ at 800 V.

**Figure 6.** Sum normalized pulse height (points) and simulated ionization loss (solid line) spectrum at 800 V.

Thus, the periodical structure in coordinate spectrum in figure 7 is almost disappeared. Taking into account the geometry of the detector, its solid angle, and neutron flux of the source an estimation of the detector efficiency ~ 4% is obtained. Simulated efficiency for two energy thresholds is presented in table 1.

**Table 1.** Expected detector efficiency via neutron wavelength for two energy thresholds

| Threshold, (MeV) | $\lambda = 1.82$ Å | $\lambda = 4$ Å | $\lambda = 8$ Å | $\lambda = 16$ Å |
|---|---|---|---|---|
| 0.01 | 6.2 % | 8.8 % | 11.6 % | 15.0 % |
| 0.2 | 4.5 % | 6.3 % | 8.2 % | 10.5 % |

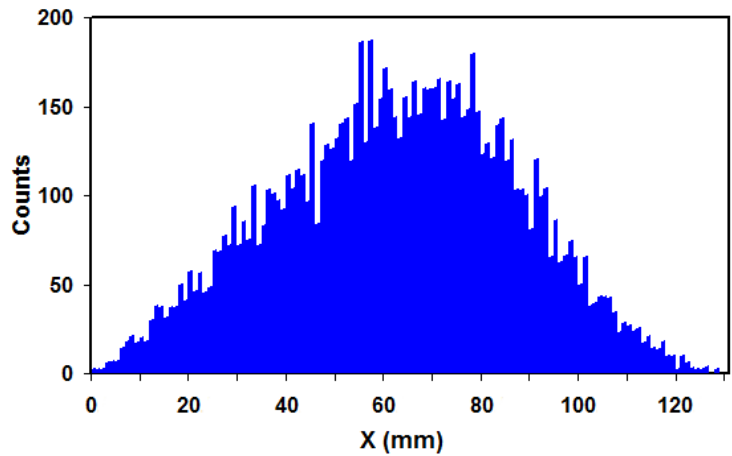

**Figure 7.** Coordinate spectrum at 800 V.

Then a cadmium shield of 2 mm thickness is installed in front of detector so that the left edge of the detector sensitive area was open but the right edge was closed on axis $X$. The opened width of detector area was 75 mm. The pulse height $X_1$ and $X_2$ 2D-diagram at 650 V is shown in figure 8. It can be seen that the main part of events (99.99%) is located inside the dashed ellipse.

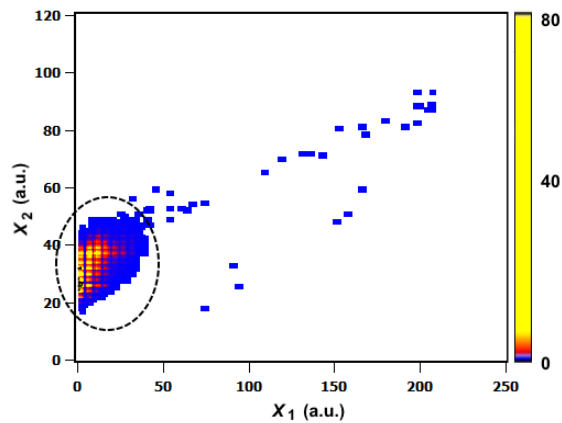

**Figure 8.** Pulse heights $X_1$ vs $X_2$ at 650 V. The dashed ellipse contains 99.99% events.

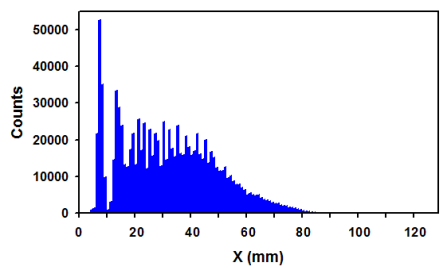

**Figure 9.** Coordinate spectrum at 650 V for events within the dashed ellipse in figure 8.

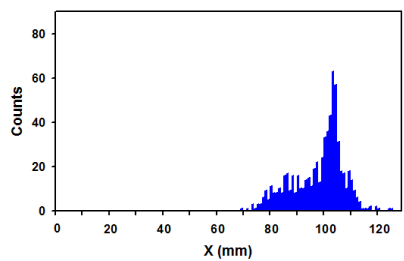

**Figure 10.** Coordinate spectrum at 650 V for events outside the dashed ellipse in figure 8.

The *X* coordinate spectrum for events located within the dashed ellipse is shown in figure 9. The shadow of the cadmium shield is observed on the right side. On the contrary, the coordinate spectrum for 0.01% events out of the dashed ellipse is shown in figure 10. They are related to the $^4$He and $^7$Li nuclei which produce a long track and move at small angle to the anode wire plane. Counting rate without Be-target (only electrons and gamma) was less than 0.001% of that with Be-target (thermal neutrons).

## 5. Conclusions

Position-sensitive thermal and cold neutron detector with 3 μm sensitive $^{10}$B layer and 128 x 128 mm$^2$ gas chamber was created and studied. Positions are determined by a division charge method. The detector efficiency is estimated as 4% to 8%. Ratio of background efficiency to thermal neutron efficiency is less than 10$^{-5}$. Pulse height resolution is about 15% and *X* coordinate spatial resolution is estimated as 2.5 mm at 700 V for gas mixture Ar+25%CO$_2$+0.3%CF$_3$Br in standard conditions.